\documentclass[twocolumn,secnumarabic,amssymb, amsmath, nofootinbib,tightenlines,
nobibnotes, aps, prc]{revtex4-1}

\bibpunct{[}{]}{;}{n}{}{,} 
\usepackage{float}

\usepackage{color}
\usepackage{caption}            
\usepackage{titlesec}

\usepackage{subcaption}
\usepackage{hyperref}
\usepackage{graphicx}
\usepackage{dcolumn}
\usepackage{bm}
\usepackage{epstopdf}

\pdfoutput=1
\begin{document}

\title{The optical potentials  and nuclear reaction cross sections for the $n$-$ ^{12}$C  and $N$-$ ^{12}$C scattering}

\author{Imane Moumene $^{1,}$ }  \email[]{Present address: Dipartimento di Fisica, Universit\`a degli Studi di Milano, Via G. Celoria 16, 20133 Milano, Italy. imane.moumene@mi.infn.it } \author{Angela Bonaccorso $^{2,}$ }
\email[]{bonac@df.unipi.it}



\affiliation{$^1$Istituto Nazionale di Fisica Nucleare, Galileo Galilei Institute for Theoretical Physics, Largo Enrico Fermi, 2, 50125 Firenze, Italy. \\     $^2$Istituto Nazionale di Fisica Nucleare, Sezione di Pisa, Largo Bruno Pontecorvo 3, 56127 Pisa, Italy. }


\date{\today}

\begin{abstract}

In this work we extend a previously derived $n$- $^9$Be optical potential  up to 500 MeV and apply it to the system $n$- $^{12}$C, finding excellent results for the energy dependence of the  total cross sections. Results obtained with a standard optical model calculation are compared to those from the eikonal formalism in order to asses the accuracy of the latter as a function of the nucleon incident energy. For comparison, single folded  (s.f.) nucleon-target potentials are also  obtained using  $^{12}$C densities from different models. These potentials are sensitive  to the density used and  none of them reproduce the characteristics of the phenomenological potential nor the cross section results.  We then calculate  nucleus-nucleus ($NN$) potentials and total reaction cross sections for some "normal"  and exotic projectile  nuclei on   $ ^ {12} $C   within the eikonal formalism. We  find that  single folded (S.F.) projectile-target imaginary potentials and  double folded  (D.F.) potentials can produce similar  energy dependence of the reaction cross sections  but the S.F. results agree better with experimental data provided  the radius parameter of the phenomenological $n$-target potential is allowed to be energy dependent. We conclude that the results previously obtained for  a  $^9$Be target are quite general, at least for light systems, and that a S.F. $NN$ potential built on a phenomenological nN potential can constitute an interesting  and useful alternative to D.F. potentials. 
\end{abstract}


\maketitle


\section{Introduction}
 
 Since its first introduction in 1958 \cite{fesh}, the optical-model
potential has been widely used to describe scattering of nucleons
and composite particles off nuclei.
As shown in Ref. \cite{fesh, fesh1}, the optical-model potential is the
single-particle operator which, introduced in the one-body Schr\"odinger
equation, yields the elastic part of the full many-channel wave
function. As Feshbach already pointed out in \cite{fesh1} the
"generalized optical-model potential" is complex, nonlocal, and
energy dependent, therefore it is very difficult to calculate without
the introduction of several approximations.
 The first-order term of the  Feshbach potential is real and it assumes a straightforward "folding" form in terms of the projectile and target densities and a nucleon-nucleon interaction. Such a form is called double folding  (D.F.) when both projectile and target are composite nuclei, while in the case of a projectile given by a single nucleon one talks of single folding  (s.f.) because only the target density is folded with the nucleon-nucleon interaction. The second-order term is complex. Its real part represents a correction to the first-order term often referred to as "polarization potential." The imaginary part represents all possible reactions between projectile and target and it is obviously difficult to calculate microscopically.

However, the O.P. (optical
potential) has been successfully applied in the framework of
a phenomenological approach in which its form factor has been chosen
on the grounds of nuclear structure considerations and its parameters
have been adjusted in order to fit the experimental data.
In spite of its complexity, several attempts have been made to
calculate the optical potential. In the past they were  mainly
concerned with the calculation of the real part of the O.P. via folding procedures, while the imaginary part has been treated phenomenologically
due to its further complexity.
 
The folding procedure which is exact for the first-order real term was  generalized by several authors \cite{3,4,SL,S2} to obtain both
the real and the imaginary part of the optical potential, 
introducing 
 an effective, complex  $g$ matrix which describes the nucleon-nucleon
interaction.
In the high energy limit one can easily obtain the Glauber \cite{59}  form of the reaction cross section in terms of the imaginary potential as given  by the folding form \cite{thesis}, which was first used by De Vries and Peng \cite{dvp} and Kox et al.\cite{kox}.

However, from the time of the introduction of folding potentials   Satchler \cite{S2} suggested that caution should be taken with the model, in particular when applied to obtain the imaginary part of the optical potential. The imaginary potential should be all orders in the interaction while the folding procedure provides first-order potentials. Furthermore a known drawback of imaginary folded potentials is that they are often too absorptive in the internal part while being too shallow on the surface. This can be a problem, for example for  exotic nuclei which are often very diffuse due to the anomalous $N/Z$ ratios and present phenomena such as neutron halo and neutron skin.
In order to improve the calculations of $NN$   folded potentials  Satchler and Love \cite{SL} proposed a different type of single folded (S.F.) potentials obtained by folding a  phenomenological nucleon-nucleus interaction with the density of  the other colliding nucleus.   The authors of Ref.\cite{sfpang} applied this idea by using  the Bruy\`eres Jeukenne-Lejeune-Mahaux
(JLMB) model  \cite{jlm,jlm1} for the $nN$ potentials  folded with various projectile densities. Recently the authors of  \cite{jin}  folded the KD02 global nucleon-target potential  \cite{KD} with $^{6,7}$Li densities. Another method  called MOL \cite{suz}, for modified optical limit,  can also be interpreted as a special kind of the S.F.  procedure that we will discuss with Eq. (\ref{5}). In Ref.\cite{suz} an effective $nN$ profile function was introduced within the Glauber approach, which acts  as the $nN$ optical potential does in the S.F.  model. 

A simple use and check of the  imaginary folded potential is in the calculation of reaction cross sections. In the past, a very detailed study of the dependence of reaction cross section values on the parameters of the folded potential was done in the seminal paper Ref. \cite{Huss}, while Ref. \cite{carlos} dealt with Pauli blocking and medium effects in nucleon knockout. In general D.F. potentials  need to be corrected to take into account  medium  effects beyond the simple $nn$ interaction.
Toward this goal, more recently,  in studying the energy dependence of reaction cross sections  by the MOL \cite{suz} Glauber approach, several groups have tried to modify some of the  ingredients of  the double folding model in the attempt to improve its performances. For example in Ref. \cite{take} the average neutron-proton $(np)$ and proton-proton $(pp)$ cross sections were  modified, while in Ref. \cite{fitbeta} the range parameters $\beta$ of the effective $(nn)$ and $(np)$ interactions were fitted.  See Eqs. (\ref{9}-{\ref11}).

More fundamental, microscopic approaches  to calculate  the imaginary potential have started to be quite successful in the  last twenty years thanks also to improved computing methods.  See Ref. \cite{bobwim} for a recent, exhaustive review. For nucleon-nucleus $(nN)$ potentials {ab initio} methods have reached a quite high degree of accuracy \cite{ch,carlo, Vorabbi1,n4lo,Vorabbi}. On the other hand  the $nN$ potential \cite{furu} and $NN$ potentials
 (Refs. \cite{Sakura,Sakura1,Sakura100,ogata} and other works by the same authors) are based on a microscopic, complex $g$ matrix and then either a single folding  or a double folding model is constructed. In the following we will define and discuss further these approaches.

However, when, for a give nucleus, a large set of data is  available it might be useful to start  fitting the parameters of a phenomenological potential. For example in Ref. \cite{bobme}, thanks to the existence of  an almost continuous series of neutron- $^9$Be  data  as a function of the neutron incident energy,   a phenomenological potential and a dispersive optical model (DOM) \cite{MS}  potential  were introduced for the system neutron- $^9$Be, and were able to reproduce at the same time the total, elastic,  and reaction cross sections and all available elastic scattering angular distributions. These results were important because they showed that  a phenomenological  nucleon-target O.P. could be obtained also for light nuclei and on a wide energy range. Then using one of those  potentials, AB, a S.F.  (light)-nucleus- $ ^9$Be imaginary optical potential was derived and it was shown that it is more accurate than a   D.F. optical potential \cite{noi1,noi2,imane} in reproducing $NN$ reaction cross section. Considering that $ ^9$Be is one of the most used targets for  a large number of reaction studies, the above cited works constituted an important starting point for further studies and applications, in particular for reactions with exotic nuclei.

 Of course one might wonder whether such results are due to the special nature of $ ^ 9$Be, which is itself weakly bound and strongly deformed. For this reason and to draw more general conclusions we decided to try to apply in this work the same AB potential to the description of $n$-$^{12}$C  scattering and calculate by the optical model total  cross sections in the range 20-500 MeV.  At the moment we do not attempt to fit the low energy resonance region, which would need an {\it ad hoc} study in particular as far as the spin-orbit potential is concerned. One motivation is that we are eventually interested in experiments with exotic nuclei studied at energies larger than about 60-80$A$. MeV. These are insensitive to the low energy part on the nucleon-target interaction, while there is a large bulk of data at relativistic energies larger than 200$A$. MeV.  For example the BARB experiment at GSI deals with high energy beams \cite{GSI}. For this reason we have extended the AB potential to fit $n$- $^9$Be and $n$- $^{12}$C  total cross sections above 200$A$. MeV, finding small differences in the two cases.  Folding the newly established $n$- $^ {12}$C optical potential with several projectile densities, we will then construct  S.F.  $N$- $^{12}$C  potentials. These potentials are necessary to calculate reaction cross section and deduce from data unknown nuclear densities and rms radii, as mentioned  above. 
Optical potentials are also necessary in breakup models to calculate the $S$ matrices for the core-target and nucleon-target  scattering.   In the future it would be interesting to apply the S.F. and D.F. potentials to a series of exotic nuclei  knockout induced reactions  in order to asses their accuracy in reproducing  single nucleon  breakup absolute cross sections as suggested in \cite{filo}. 

Besides  fundamental research, we would like to stress the other important application of    simple optical potentials in transport codes. An essential ingredient of such codes are  the calculated realistic nuclear reaction cross sections used for risk evaluation of manned space exploration missions as
well as for radiation therapy, where one needs dose calculations for treatment planning \cite{luoni}.   The
therapeutic use of heavy ions, such as carbon, has gained significant interest due to
advantageous physical and radiobiologic properties compared to photon based therapy \cite{pet}. Most recently exotic nuclei close to $^{12}$C, such as $^{12}$N, $^{11}$C,  and $^{10}$C have been proposed  for radiation therapy \cite{GSI}. Also in reactor physics data and models of
reaction cross sections are of fundamental importance \cite{kuni}. 


 Turning to theoretical methods  which use the O.P. to obtain  reaction cross sections, while the optical model (OM) and coupled-channel (CC) model are certainly the most accurate ways, as  previously mentioned, the Glauber model \cite{59} with folded potentials  (f.p.) \cite{SL,S2}, has also been used for many years \cite{dvp,kox} and its results have been compared to  data. In particular from the beginning of physics with  radioactive ion beams (RIBs)  the method has become very popular for its simplicity in  deducing density distributions of exotic nuclei and their root mean square  (rms) radii \cite{tanih1,tanih,ozawa,Ca+12C,fitbeta,take,hor1,hor2,suz} and the core-target survival probability in knockout reactions \cite{AB0}.  In particular in a recent work \cite{filo}  the sensitivity to folding methods used to obtain the nucleon-target and core-target optical potentials in standard knockout eikonal calculations  used to  extract spectroscopic factors has been discussed.

  A brief reminder of basic formulas  used in this paper is provided in  Sec. II. Then Sec. III, which contains our results, is divided   in two subsections. In the first  the extension of the  $n$- $^9$Be AB potential of Ref. \cite{bobme} up to 500 MeV is provided and it is shown that almost the same potential can be applied to $n$- $^{12}$C scattering. 
 Cross sections are calculated with a standard optical model and with the eikonal method using  the phenomenological potential and some folded $n$-target s.f. potentials. Our focus will be on the comparison of  results for the energy dependence of the total cross sections. In this way, for the $n$-target system,  we will test the accuracy of the phenomenological potential vs the s.f. potential, the dependence on the  target model density and of the optical model vs the Glauber model.
To lend further support to our  S.F. approach,  similarly to what has been done in Refs. \cite{noi1, noi2}, we will calculate in   Sec. \ref{IIIB} , the imaginary part of  $^{12}$C- $^{ 12}$C optical potential  with the S.F. potential  built from the projectile density and the phenomenological $n$-target potential, and with the  D.F. potential obtained from the projectile and target densities and the nucleon-nucleon interaction, and discuss their differences. Finally   $NN$ reaction cross section calculations made with the two different potentials  will be compared to experimental values for the systems  $^{12}$C+ $^{ 12}$C, $^{9}$Be+ $^{ 12}$C,  $^{20}$Ne+ $^{ 12}$C, $^{n}$Ca+ $^{ 12}$C. Given the symmetry of projectile and target, the first system is a particularly interesting test case for the accuracy of the  phenomenological potential approach vs folded potential.
   
For the various  cases studied, we will provide figures of the radial dependence of the imaginary potentials used,  their volume integrals, and rms radii, such that differences in cross section results can be traced back to how various potentials represent the localization of reactions and on how they might contain  {\it in-medium} and short-range repulsion effects.

\section{Theory}  The $n$- $^9 $Be phenomenological potential AB of Ref. \cite{bobme}  is here extended to 500 MeV and to the system n+ $^{ 12}$C. The potential of this paper has the form
   
   \begin{equation}
U_{AB} (r,E) =-\left[V_{\text{WS}}(r,E)+iW_{\text{WS}}(r,E)\right].   \label{1} 
\end{equation}
The real part of the neutron-target interaction is given by $V_{\text{WS}}$, the usual Woods-Saxon potential:
\begin{equation}
V_{\text{WS}}(r)=V^R f(r,R^{R},a^R)\label{vr}.
\end{equation}

Also, the imaginary part takes the form
\begin{equation}
W_{\text{WS}}(r)=W^{vol} f(r,R^{I},a^I)-4 a^{I} W^{sur} \frac{d}{dr} f(r,R^{I},a^{I}). \label{3}\end{equation}
with $f(r,R^{i},a^i)=\left (1+e^{\frac{r-R^{i}}{a^{i}}}\right)^{-1} $ and 
$
R^i = r^i A^{1/3}
$.

The real AB potential of Ref. \cite{bobme} contained also a correction term  $\delta V$  which originates from surface-deformation effects and represents channels for  which a simple Woods-Saxon form is not appropriate. Because such couplings are important only up to around 20 MeV, and  here we are not interested in this low energy region  for the present applications on $^{ 12}$C, we shall take   $\delta V  =0$. For the same reason the spin-orbit term will be neglected. The parameters of $U_{\text{AB}}(r,E)$
for the \textit{n}- $^{ 9}$Be and \textit{n}- $^{12}$C interaction used in this paper are given in Tables I and II respectively.

  \squeezetable
\begin{table*} []
\caption{Energy-dependent optical-model parameters for the (AB) potential for $n+ ^{9}$Be. $r^I = 1.3 $ fm, $a^I = 0.3$ fm at all energies. See also Table \ref{tabri} and text.}
\label{tab1} 
\begin{ruledtabular}
\begin{tabular} {cccccc}  
$E_{\text{lab}}$&$V^R$&$r^R$&$a^R$&$W^{sur}$&
$W^{vol}$\\
(MeV)&(MeV)&(fm)&(fm)&(MeV)&(MeV)\\&&&&&\\
\hline
20$\leq E_{\text{lab}}<$40&$31.304-0.145E_{\text{lab}}$&$1.647-0.005(E_{\text{lab}}-5)$&0.3-0.0001$E_{\text{lab}}$&$1.65+0.365E_{\text{lab}}$&$5.6-0.005(E_{\text{lab}}-20)$\\    
40$\leq E_{\text{lab}}<$111&''&''&''&$16.25-0.05(E_{\text{lab}}-40)$&$5.5-0.01(E_{\text{lab}}-40)$\\ 
111$\leq E_{\text{lab}}<$160&''&''&0.288&12.7&4.8\\ 
		$160 \leq E_{\text{lab}}<$200&''&''&''&$12.7-0.025(E_{\text{lab}}-160)$&$4.8-0.025(E_{\text{lab}}-160)$\\
		$200\leq E_{\text{lab}}<$215&"&"&"&$11.7+0.02(E_{\text{lab}}-200)$&$3.8+0.02(E_{\text{lab}}-200)$\\
		$ 215 \leq E_{\text{lab}}\leq $500&0&"&"&"&"  
\end{tabular}
\end{ruledtabular}
   \end{table*}
  \squeezetable
 \begin{table*} []
 	\caption{Energy-dependent optical-model parameters of the potential $n$- $^{ 12}$C for  $E_\text{{\text{lab}}}\ge 160$ MeV.  At lower energies, the parametrization is the same as for  $^{9 }$Be in Table \ref{tab1}.}
 	\label{tab2}
\begin{ruledtabular}
\begin{tabular} {cccccc}  
$E_{\text{lab}}$&$V^R$&$r^R$&$a^R$&$W^{sur}$&
$W^{vol}$\\
(MeV)&(MeV)&(fm)&(fm)&(MeV)&(MeV)\\&&&&&\\
\hline
$160 \leq E_{\text{lab}}<$200&$31.304-0.145E_{\text{lab}}$&$1.647-0.005(E_{\text{lab}}-5)$&0.288&$12.7-0.025(E_{\text{lab}}-160)$&$4.8-0.025(E_{\text{lab}}-160)$\\
$200\leq E_{\text{lab}}<$215&"&"&"&11.7&3.8\\
$ 215 \leq E_{\text{lab}}< $220&0&"&"&"&"\\
$ 220 \leq E_{\text{lab}}\leq$ 500&"&0.1&"&$11.7+0.02(E_{\text{lab}}-220)$&$3.8+0.02(E_{\text{lab}}-220)$    
\end{tabular}
\end{ruledtabular}
  \end{table*}

\begin{table}[ht]
	 \caption{Energy-dependent optical-model parameter $r^I$ for the (AB) potential for $n+^ { 9}$Be and $n+ ^{ 12}$C used in calculations of S.F. $NN$ potentials.}
	\begin{center}		
	{\renewcommand{\arraystretch}{1.5}
			{\setlength{\tabcolsep}{.3cm} 
\begin{tabular} {ccc}  
\hline\hline
$E_{\text{lab}}$&$r^{I}(^9$Be)&$r^{I}(^{12}$C)\\
(MeV)&(fm)&(fm)\\&&\\

\hline
 30$\leq E_{\text{lab}}\leq $160&$1.4- 0.0015E_{\text{lab}}$&$1.32- 0.0013E_{\text{lab}}$\\    
 $E_{\text{lab}}>$160&1.15&1.118\\ 
 \hline\hline
\end{tabular}}}

 \label{tabri}
	\end{center}
	\end{table}

For comparison we consider also  a s.f. \cite{S2,thesis}  $n$-target potential $U^{nT}_{\rho}$  defined as

  \begin{eqnarray}
U^{nT}_{\rho}({\bf r})=-{1\over 2} \hbar v \sigma_{nn}(1 -i\alpha_{nn} )\rho_T ({\bf r})\label{4}, \end{eqnarray}
where $\rho_T ({\bf r})$ is the target density function, for which we will use a number of different models as specified in the following, $\sigma_{nn}$ is the average of the experimental neutron-proton and proton-proton cross sections, and  $\alpha_{nn}$ is the ratio of the real and imaginary scattering amplitude at zero degrees. $v$ is the classical relative motion velocity of the scattering.   The previous equation can be generalized in an obvious way in order to distinguish between the proton and neutron densities and the proton-neutron and proton-proton cross sections, using
$\rho_P={\rho^n}_P+{\rho^p}_P$ and $ U^{nT}_{\rho}({r})=-{1\over 2} \hbar v [\sigma_{np}(1-i\alpha_{np}){\rho^p}_T ({r})+ \sigma_{pp}(1-i\alpha_{pp}){\rho^n}_T ({r})]$. This is the formalism followed in the present work. Here we are assuming a zero-range nucleon-nucleon interaction,  and in numerical calculations the values of  $\sigma_{nn}$  and $\alpha_{nn}$  will be taken from the parametrization of Refs. \cite{hor1, hor2, carlos}. 

In the case of $NN$ scattering we will discuss  potentials 
$U^{NN}$, negative defined as

\begin{equation}U^{NN}({\bf r})=\int  d{\bf b_1} U^{nN}({\bf b_1}-{\bf b},z) \int dz_1~\rho({\bf b_1},z_1). \label{5}\end{equation}
This quantity is the  S.F. optical potential given in terms     of a nucleon-nucleus ($nN$) optical potential $U^{nN}({\bf r})$ and the matter density $\rho({\bf b_1},z_1)$ of the other nucleus. In the S.F. method,  $U^{nN}({\bf r})$ can be a phenomenological nucleon-target potential, Eq. (\ref{1}), such as the DOM  or the AB potentials of Ref. \cite{bobme}. In the D.F. method, $U^{NN}$ is obtained  from  the microscopic densities $\rho_{P,T} ({\bf r})$ for the projectile and target respectively and an energy-dependent nucleon-nucleon ($nn$) cross section $\sigma_{nn}$, by using Eq. (\ref{4}) for $U^{nN}$ with the notation  $T = N$  in Eq. (\ref{5}).

The reaction cross section, which depends only on the imaginary potential, in the eikonal formalism is given by the well known formula 
\begin{equation}
\sigma_{R}=2\pi\int_0^{\infty}  b~d b~ (1-\mid S_{PT}({\bf b})\mid ^2) \label{6}  \end{equation} 
where
\begin{equation}\mid S_{PT}({\bf b})\mid ^2=e^{2\chi_I(b)}\label{7}\end{equation} is the probability that the projectile-target (PT) scattering is elastic for a given impact parameter $\bf b$.

 The imaginary part of the eikonal phase shift is given by \begin{eqnarray}\chi_I({\bf b})&=&-{1\over \hbar v}\int dz~ W^{PT}({\bf b},z), \label{8}\end{eqnarray}   
where, depending on the case studied, $W^\text{{PT}}$ will be the imaginary part of one of  the nucleon-target or nucleus-target potentials defined above.

\begin{figure}[ht]
	\begin{minipage}{0.5\textwidth}
		\includegraphics[width=\linewidth]{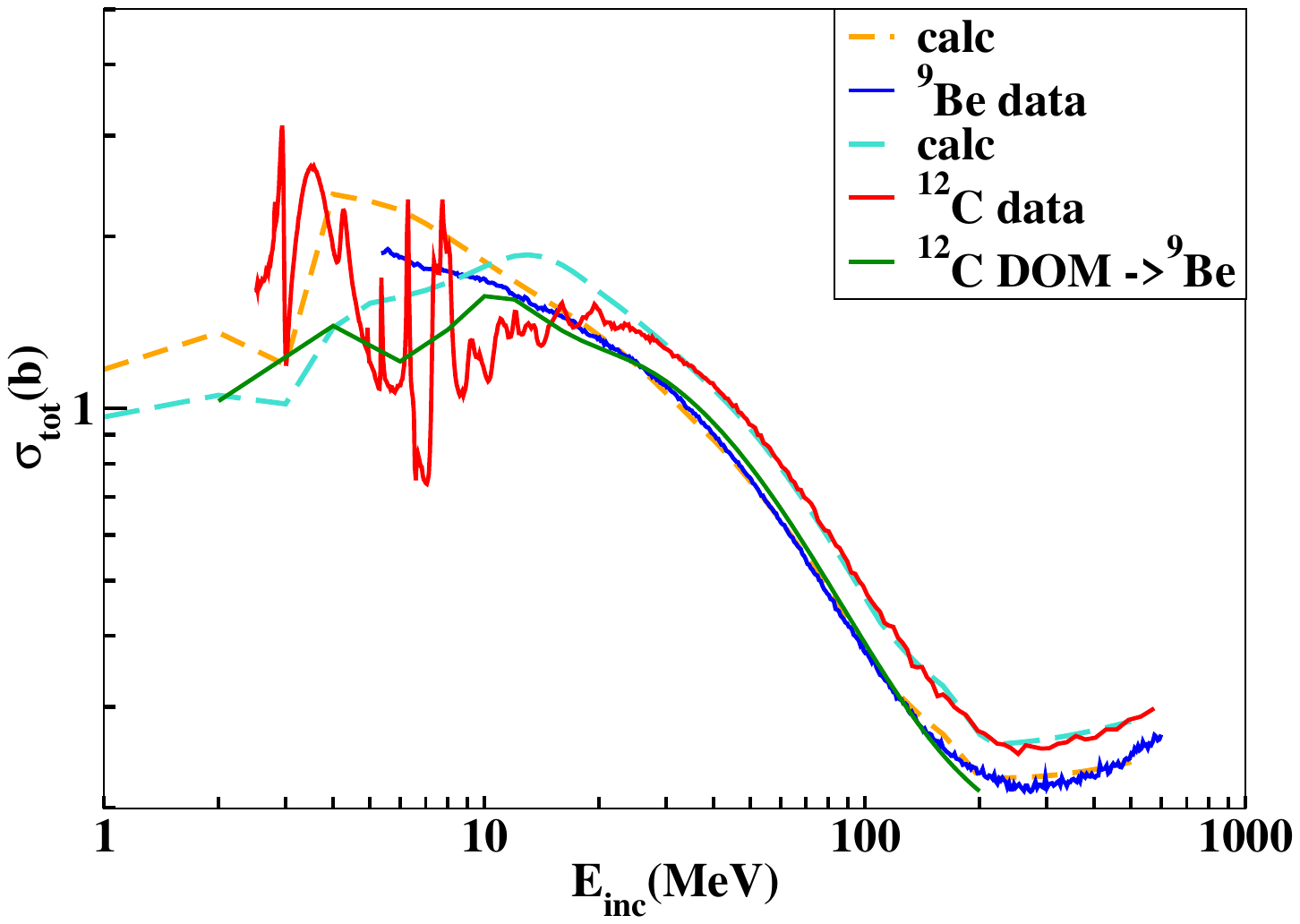}
		\captionof{figure}{(Color online) Total experimental and calculated  cross sections. Lower  blue symbols are  for $n+ ^9$Be,  upper  red symbols for $ n+  ^{12}$C. The optical model calculations are given by the orange and cyan  dashed lines, respectively. The solid  green line  is a calculation made with a DOM potential obtained for $n+ ^{12}$C and applied to $n+^{9}$Be \cite{mack}.} \label{xs1}
	\end{minipage} \hfill
	\begin{minipage}{0.5\textwidth}
		\includegraphics[width=\linewidth]{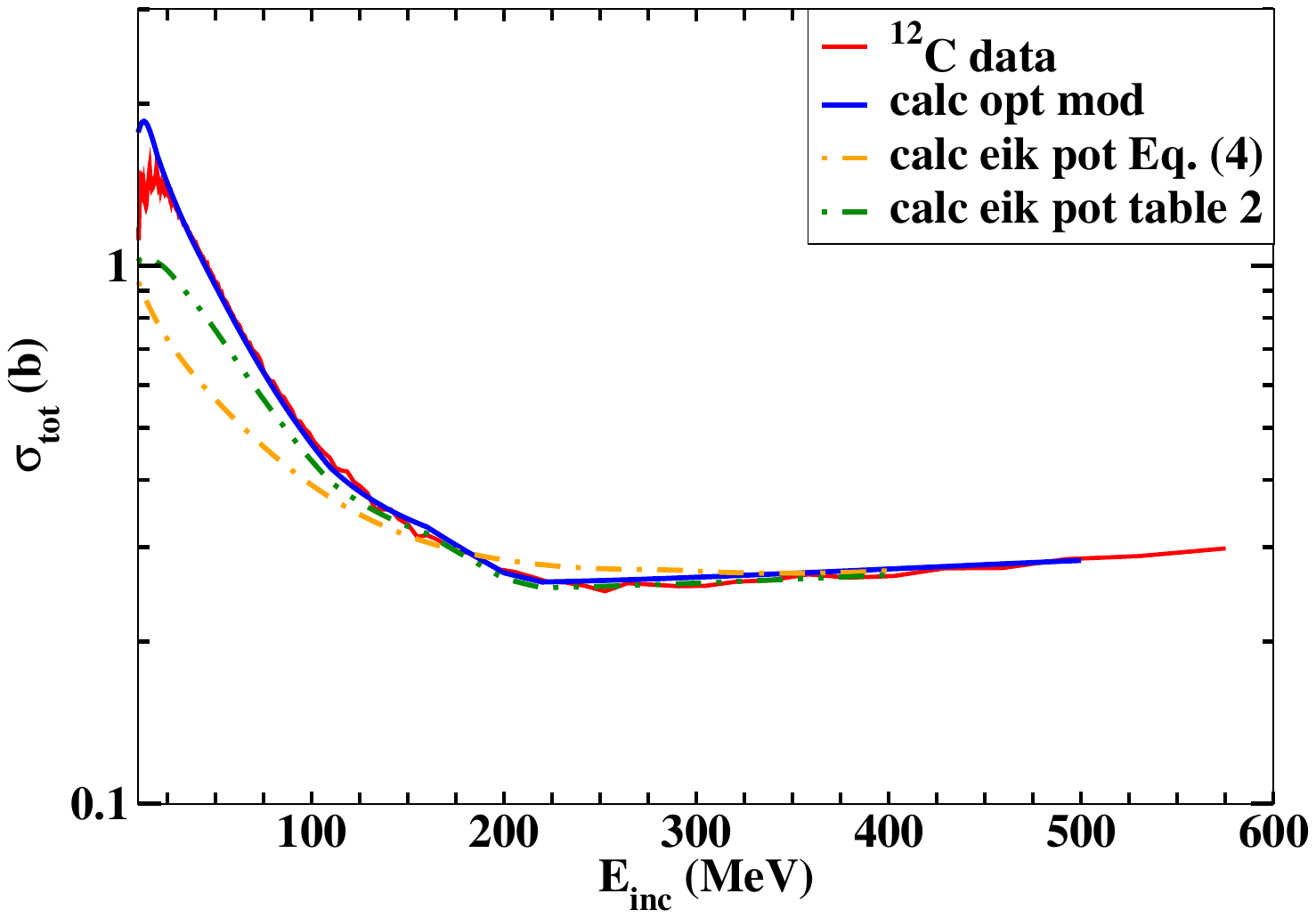}
		\hfill\captionof{figure}{(Color online) Total experimental and calculated  cross sections for  $n+  ^{12}$C. Red symbols are the data. The blue full curve and green double-dotted-dashed line are results of optical model  and eikonal calculations respectively,  with the potentials (\ref{1})-(\ref{3}) and Table \ref{tab2}. The orange dot-dashed line is the eikonal calculation with the s.f. potential  (\ref{4}).}
		
		\label{xs2}
	\end{minipage}
\end{figure}

\begin{figure}[ht]
	\begin{minipage}{0.5\textwidth}
 \includegraphics[width=\linewidth]{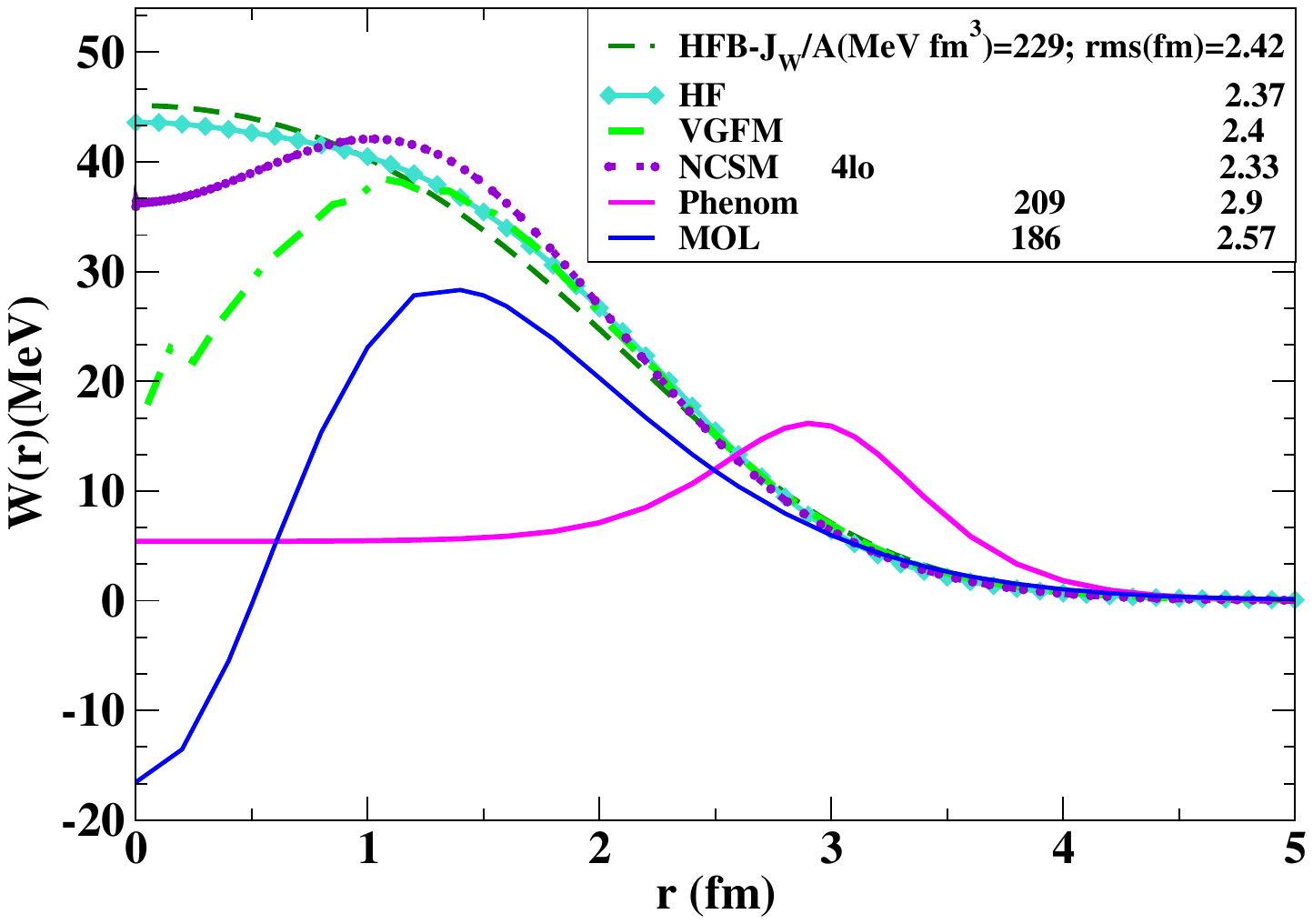}
	\caption {(Color online) $n+  ^{12}$C potentials calculated with various model densities at 300 MeV; see legend and text. The blue line is the potential deduced from the profile function of Ref. \cite{suz}. The magenta tick curve is the phenomenological potential of Eqs. (\ref{1})-(\ref{3}) and Table \ref{tab2}.}
	\label{xs4}
	\includegraphics[width=\linewidth]{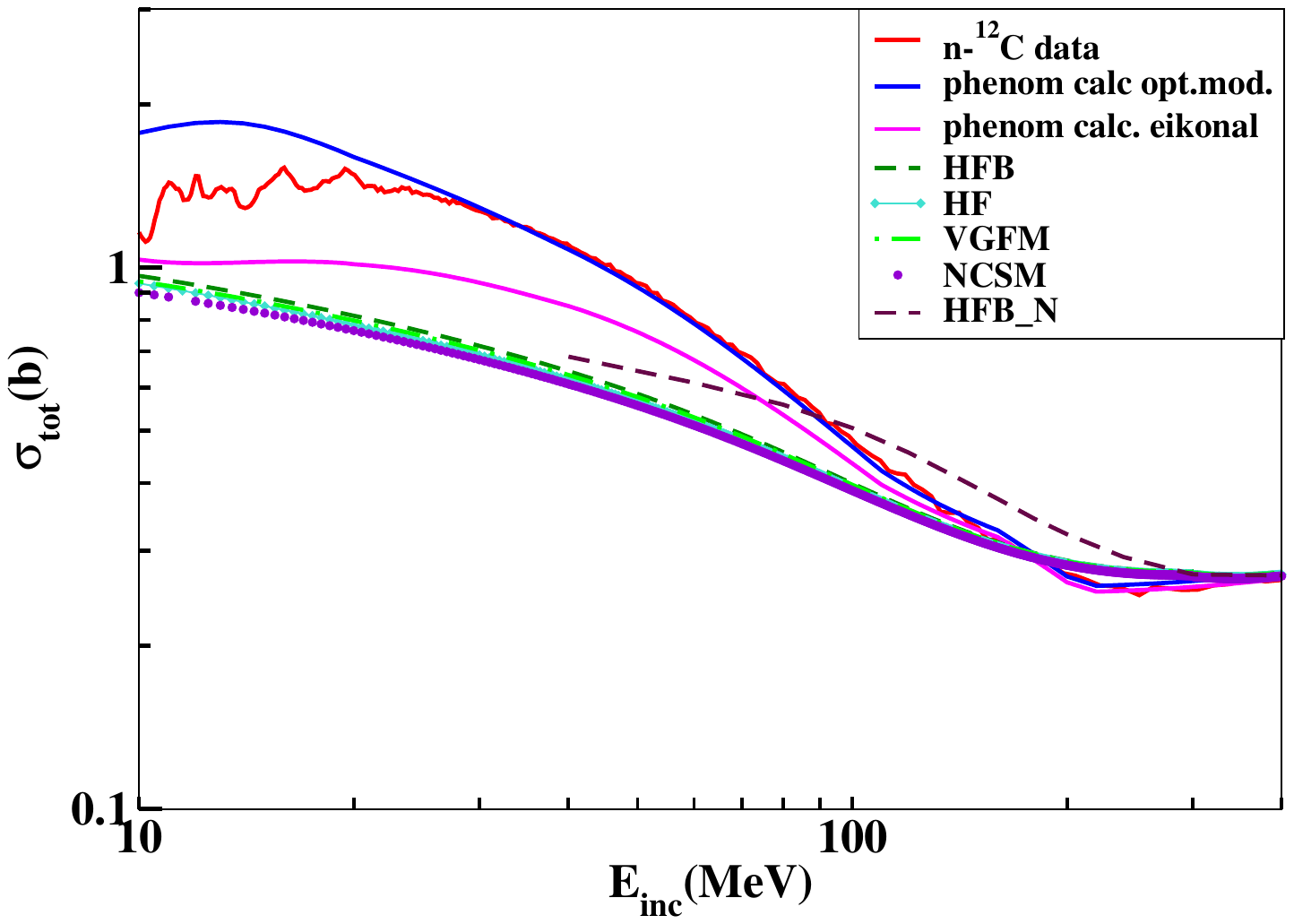}
		\caption {(Color online) Total experimental and calculated  cross sections for  $n+  ^{12}$C. Red symbols are the data.  The blue curve is the calculation by the optical model with the phenomenological potential. The other curves are calculations using the  s.f. potential  (\ref{4}) and  Fig. \ref{xs4} using fixed $\alpha_{nn}$ values in Eq. (\ref{4}) appropriate for 300 MeV. The brown dashed curve labeled as HFB\_N uses the energy dependent $\alpha_{nn}$ from Refs. \cite{hor1, hor2, carlos}. Note that they are known only from 40 MeV. See text for details.}
			\label{xs3}
	\end{minipage}
	\begin{minipage}{0.5\textwidth}
	
\end{minipage}
\end{figure}


 \squeezetable
\begin{table*} []
\begin{ruledtabular}
	{  \renewcommand{\arraystretch}{1.5}
			{\setlength{\tabcolsep}{.3cm}
			\caption {Comparison of the reaction cross sections of the $^{12}$C+$^{12}${C} system. Incident energies  are indicated  in the  first column.  Strong absorption radius parameters within the single and double folding methods are listed in the third column. The  fourth column provides the volume integrals for active particles. The next  columns contain the  theoretical cross sections calculated with various densities. Before each of them are the rms radii of the corresponding imaginary potentials, some of which are shown in Fig. \ref{pot4}.}
				\begin{tabular}{cccccccccc}		
					$E_{\text{inc}}$&Model& $r_{s}$&$J_{W}/A_PA_T$&rms radius&$\sigma_{\text{NCSM}}$ &rms radius&$\sigma_{\text{HF}}$ &rms radius&$\sigma_{\text{HFB}}$ \\	
     (MeV)&& (fm)&(MeVfm$^3$)&(fm)&(mb)& (fm)& (mb)&(fm)&(mb)\\	
					\hline 	
					83&{\scriptsize S.F.}&1.2&184&3.72&994&3.75&1008&3.78&1025\\
					&	{\scriptsize D.F.}&1.22&279&3.29&957&3.36&995&3.43&1027\\
					\hline 
					300&{\scriptsize S.F.}&1.18&151&3.57&760&3.60&768&3.64&780\\
					&	{\scriptsize D.F.}&1.11&241&3.29&791&3.36&815&3.43& 842\
					\label{TC_12C+12C}	
					\end{tabular}}}
						\end{ruledtabular}
	\end{table*}

\begin{table}[ht]
			\begin{center}
	  {\renewcommand{\arraystretch}{1.5}
			 { \setlength{\tabcolsep}{.3cm} 
				\caption{ Results for the $^{20}${Ne}+$^{12}${C} scattering. The strong absorption radius parameter is listed in the third column,  and the fourth and the fifth columns give  the predicted and the experimental \cite{kox} reaction cross sections. The  HFB density is used for $^{20}${Ne}.}
			\begin{tabular}{ccccc}		\hline
					$E_{\text{inc}}$(MeV)&Model& $r_{s}$(fm)&$\sigma_{\text{theo}}$ (mb)&$\sigma_{\text{exp}}$ (mb)\\	
					\hline 
                    30&{\scriptsize S.F.}&(1.35) 1.33&(1478) 1456&1550 $\pm  $75 \\
					&	{\scriptsize D.F.}&1.37&1560&\\
					\hline 
					100&{\scriptsize S.F.}&(1.27) 1.23&(1327)1211 &1161 $\pm  $ 80 \\
					&	{\scriptsize D.F.}&1.21&1206&\\
					\hline 
					
					200&{\scriptsize S.F.}&(1.21)1.11&(1193) 1012&1123 $\pm  $ 80 \\
		
					&	{\scriptsize D.F.}&1.15&1079&\\
					\hline 
					
					300&{\scriptsize S.F.}&(1.21)1.12&(1181)1001&1168 $\pm  $ 100\\
				
					&	{\scriptsize D.F.}&1.13&1062& \\
					\hline
					\label{20Ne+12C}
							\end{tabular}}}
																\end{center}
\end{table}

     
\begin{table}[ht]
	\caption{Results for system  $^{n}$Ca- $^{12}$C  at $E=280A$ MeV. The strong absorption radius parameter is listed in the third column,  and the fourth and the fifth columns give  the predicted and the experimental  \cite{Ca+12C} reaction cross sections. Statistical and systematic errors for the experimental values are given in the first and second parentheses  respectively. The  root-mean-square (rms) matter radius of the HFB  projectile density is listed in the last column.}
	\begin{center}
		{\renewcommand{\arraystretch}{1.5}
			{\setlength{\tabcolsep}{.15cm} 
				
				\begin{tabular}{cccccc}		
					\hline
					Nucleus&Model& $r_{s}$(fm)&$\sigma_{\text{theo}}$ (mb)&$\sigma_{\text{exp}}$ (mb)&rms  radius (fm)\\	
					\hline \hline
					$^{42}{Ca}$&{\scriptsize S.F.}&(1.23)1.14 &(1598) 1388&1463(13)(6)&3.38\\
					&	{\scriptsize D.F.}&1.16&1460&&\\
					\hline
					$^{43}{Ca}$&{\scriptsize S.F.}&(1.22)1.14&(1614)1402&1476(11)(6)& 3.40\\
					&	{\scriptsize D.F.}&1.17&1476&&\\
					\hline
					
					$^{44}{Ca}$&{\scriptsize S.F.}&(1.23)1.15&(1630) 1417&1503(12)(6) &3.42\\
					&	{\scriptsize D.F.}&1.16& 1490&&\\
					\hline
					
					$^{46}{Ca}$&{\scriptsize S.F.}&(1.24)1.15& (1683)1466&1505(8)(6)&3.50\\
					&	{\scriptsize D.F.}&1.17&1543&&\\
					\hline
					
					$^{48}{Ca}$&{\scriptsize S.F.}&(1.23)1.16& (1714)1495&1498(17)(6)&3.50\\
					&	{\scriptsize D.F.}&1.18&1573  &&\\
					\hline

		\end{tabular}}}
	\end{center}	
	\label{Ca+12C}
\end{table}

\section{Results} 

\subsection{Nucleon-$^{12}$C}
We start by showing in Fig. \ref{xs1}  the energy dependence of the total cross section calculated with an optical model code using the potential defined by Eqs. (\ref{1})-(\ref{3}) and the parameters given in Tables \ref{tab1}   and  \ref{tab2}  for $n+  ^9$Be and $n+  ^{12}$C. We include also the experimental data from Ref. \cite{exf}. It is interesting that  the experimental data exhibit a clear scaling between the two nuclei, which  the calculations reproduce accurately. Note that  the two corresponding potentials have the same radius parameter but different radii, due to the difference in mass. Otherwise the other parameters  differ only  above 160 MeV. Reference \cite{bobme}  presented also results for  $n+ ^{9}$Be from a dispersive optical potential DOM calculation.  DOM potentials exist also for   $n+ ^{12}$C. Indeed in the same figure the green solid line shows the results obtained for  a $ ^{9}$Be target using the DOM obtained for $ ^{12}$C \cite{mack}. It is amazing that, also for the DOM potential model,  the same parametrization can be successfully applied to the two different targets. As  was found in Ref. \cite{bobme} for $ ^9$Be, the agreement shown here for the $ ^{12}$C target, between data and OM calculations, is remarkable and is comparable to that obtained for example in Ref. \cite{kuni},  where a coupled-channel (CC) technique was used. Note that also the authors of Ref. \cite{kuni} stressed a similarity between parametrizations for $ ^{9}$Be and $ ^{12}$C. As we shall see in the following, the advantage of a simple OP approach, with respect to CC calculations, is that it can easily be used to build folding potentials for nucleus-nucleus scattering and also it can be used in eikonal and fully quantum-mechanical models \cite{AB0,jinme} of knockout from exotic nuclei.
 
 In Fig. \ref{xs2} the total experimental  cross section for  $n+  ^{12}$C is shown again by red symbols while the  blue full curve and green double-dotted-dashed line are results of the optical model  and eikonal calculations, Ref.\cite {59}, respectively,  with the potential of Eqs. (\ref{1})-(\ref{3}) and Table \ref{tab2}. The orange dot-dashed line is the eikonal calculation with the s.f. potential  (\ref{4}). These results indicate that, while the simple eikonal approximation with the phenomenological potential works well from about 100 MeV incident energy, the eikonal  model with the folded potential  starts to work well only from about 200 MeV. Clearly the Glauber and folding models miss some effects  of excitation modes in the target, beyond the simple $nn$ free scattering concept. The optical model with the phenomenological $n$-T potential includes instead such effects. In this respect, we first
note that the $U^{nT}_{\rho}$ potential  of Eq. (\ref{4}) has the same range and profile as the target
density because
$\sigma_{nn}$  and $\alpha_{nn}$ are simple scaling factors. To understand better this point  Fig. \ref{xs4} shows the imaginary potentials calculated at 300 MeV  with the densities indicated in the legend from Refs.  \cite{Wiringa}, \cite{nav}. Hartree-Fock-Bogoliubov (HFB) densities were calculated with the code HFBTHO \cite{HFB} and the Skyrme interaction  SkM* \cite{SkM*}. Using other Skyrme interactions  does not produce substantial differences. No-core-shell-model   (NCSM) densities  were obtained by using the nn4lo \cite{n4lo} interaction.  We provide also the volume integrals per particle and rms radius values. The former ($J_W/A_T$) have all the same values because all densities are normalized to the number of nucleons. The latter (rms values) have very similar values although in the internal parts the potentials are quite different. The phenomenological potential is completely different, being very shallow at the interior and having instead a pronounced surface peak and long tail. Its volume integral is smaller than that of the s.f. potentials while its rms radius  is much larger. Indeed Fig. \ref{xs3} shows again the experimental cross sections as in Figs. \ref{xs1} and \ref{xs2} but this time, besides the optical model calculation with the phenomenological potential, results are shown of the eikonal approximation Ref.\cite {59} with the s.f. potentials (\ref{4}) of Fig. \ref{xs4} obtained with different densities. One can notice the small effect of changing the target  density. However, it is interesting to note that the cross section values seem to scale with the rms radius  of the potential.  This result  suggests that only the surface behavior of the potential (and of the target density) determine the value of the cross section, and  that in turn it is only the rms radius of the target density that can be deduced from data, a confirmation of the simple geometrical nature of the Glauber model. In this figure the calculations marked  as  HFB\_N were made from 40 MeV using the HFB density and $\sigma_{nn}$  and $\alpha_{nn}$   taken from the parametrization of Refs. \cite{hor1, hor2, carlos} (brown dashed curve), while in the other calculations with various densities we kept   $\alpha_{nn}$ fixed at the value appropriate to 300 MeV just to show the small dependence on the density. Note that a precise evaluation of the $\alpha_{nn}$ parameters is  a delicate issue which to our knowledge has not been fully resolved to date; see in particular Fig. 4 of \cite{sch}.

\subsection{Nucleus-$^{12}$C}
\label{IIIB}
We turn now to the study of nucleus-nucleus scattering by building a D.F. potential   and a S.F. potential according to Eq. (\ref{5}). Note that   s.f. refers to a potential  for  $n$-T scattering, built on the target density, Eq. (\ref{4}), while in the case of $NN$ scattering S.F. indicates a potential built  using in Eq. (\ref{5})  the projectile density and  the $n$-T phenomenological potential, Eq. (\ref{1}). D.F. refers to a $NN$ potential obtained using Eq. (\ref{4}) in Eq. (\ref{5}).

\begin{figure}[ht]
	\begin{minipage}{0.5\textwidth}
 \includegraphics[width=\linewidth]{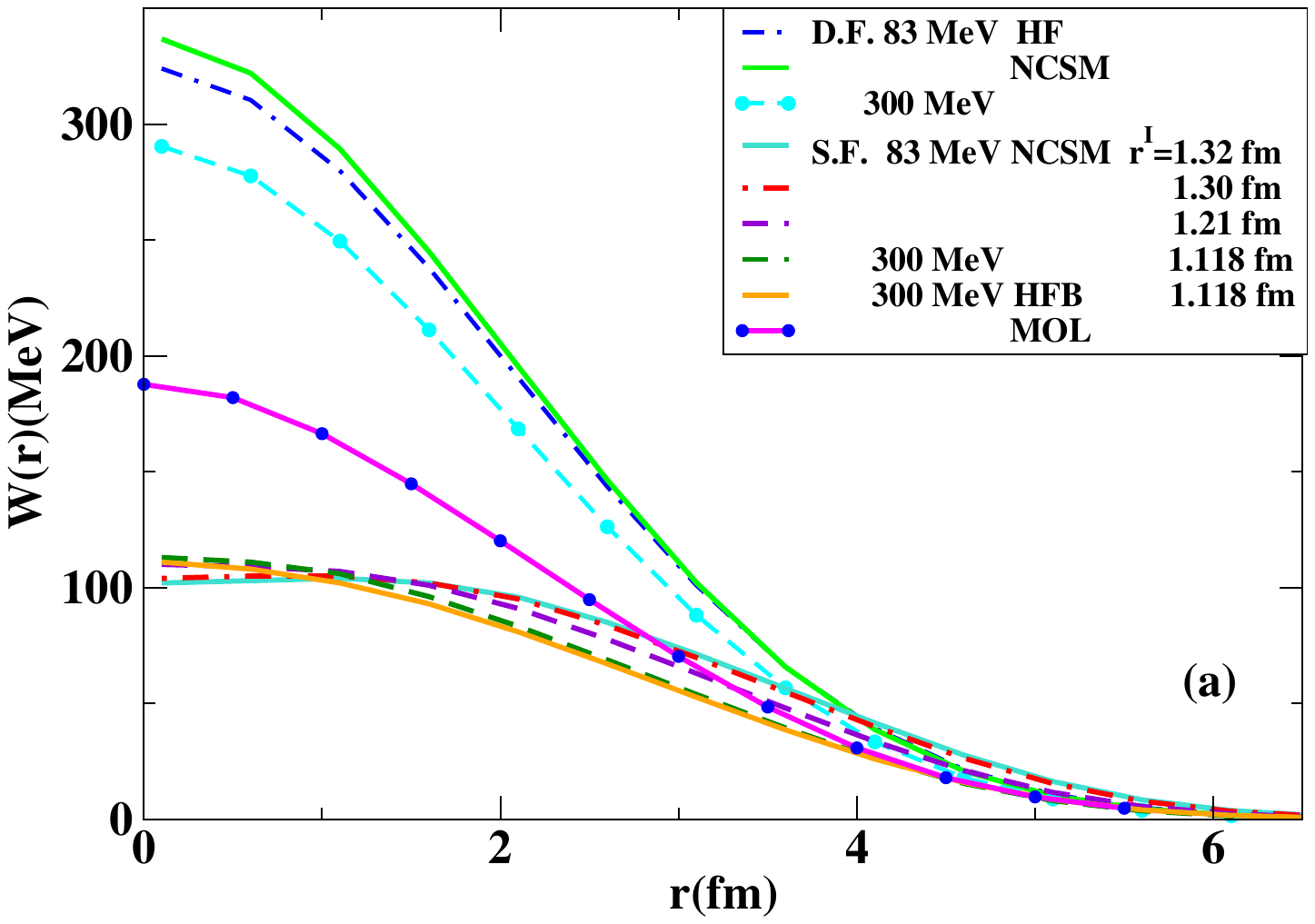}
  \includegraphics[width=\linewidth]{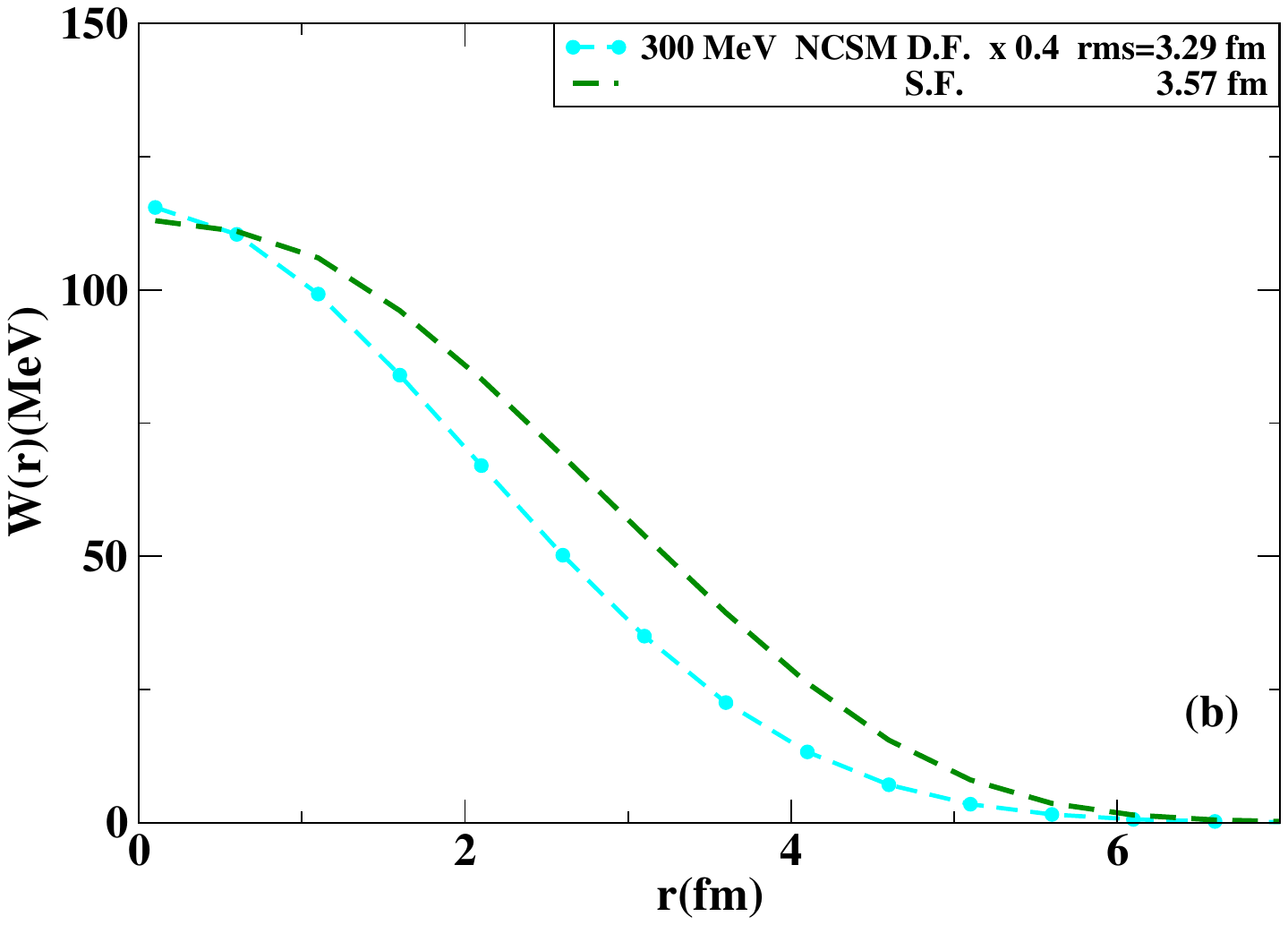}
		\caption {(Color online) (a) Imaginary part of the $^{12}$C- $^{12}$C optical potential at 83 and 300 MeV as indicated in the legend. The D.F. potentials shown are obtained with the HF and NCSM densities. The  S.F. potentials are obtained with the potentials of Table II varying the $r^I$ values and the NCSM and HFB  densities. See text for details. The full magenta line with blue uses the MOL potential obtained from \cite{suz}. Panel (b)  contains the potentials from the NCSM density at 300$A$ MeV where the D.F. has been renormalized by a factor 0.4 in order to emphasize the difference in shape and rms radius.}
			\label{pot4}
	\end{minipage}
	\begin{minipage}{0.5\textwidth}
	
\end{minipage}
\end{figure}

\begin{figure}[ht]
\begin{subfigure}{.5\textwidth}
	\centering\includegraphics[width=\linewidth]{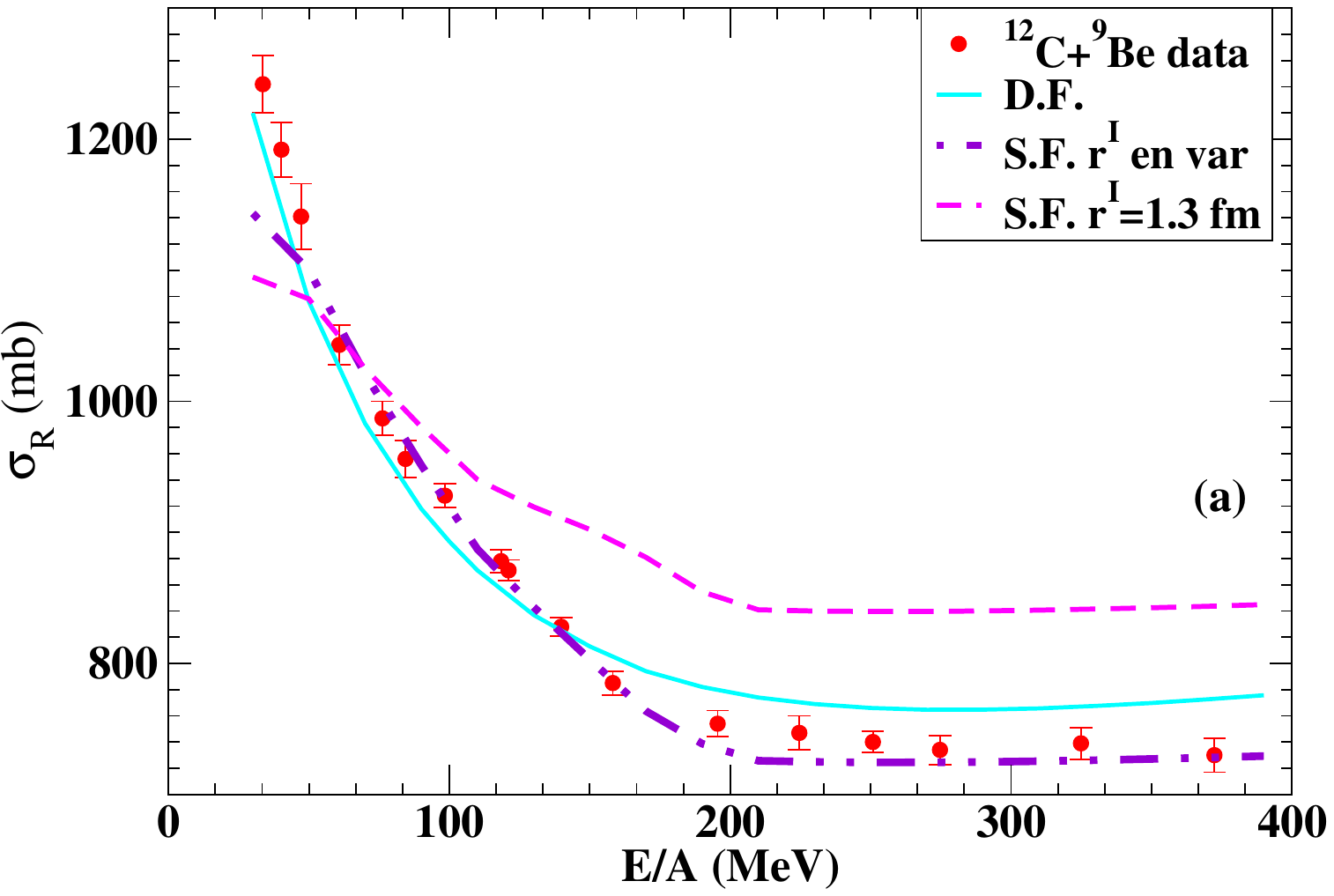}
\end{subfigure}
\begin{subfigure}{.5\textwidth}
	\centering	\includegraphics[width=\linewidth]{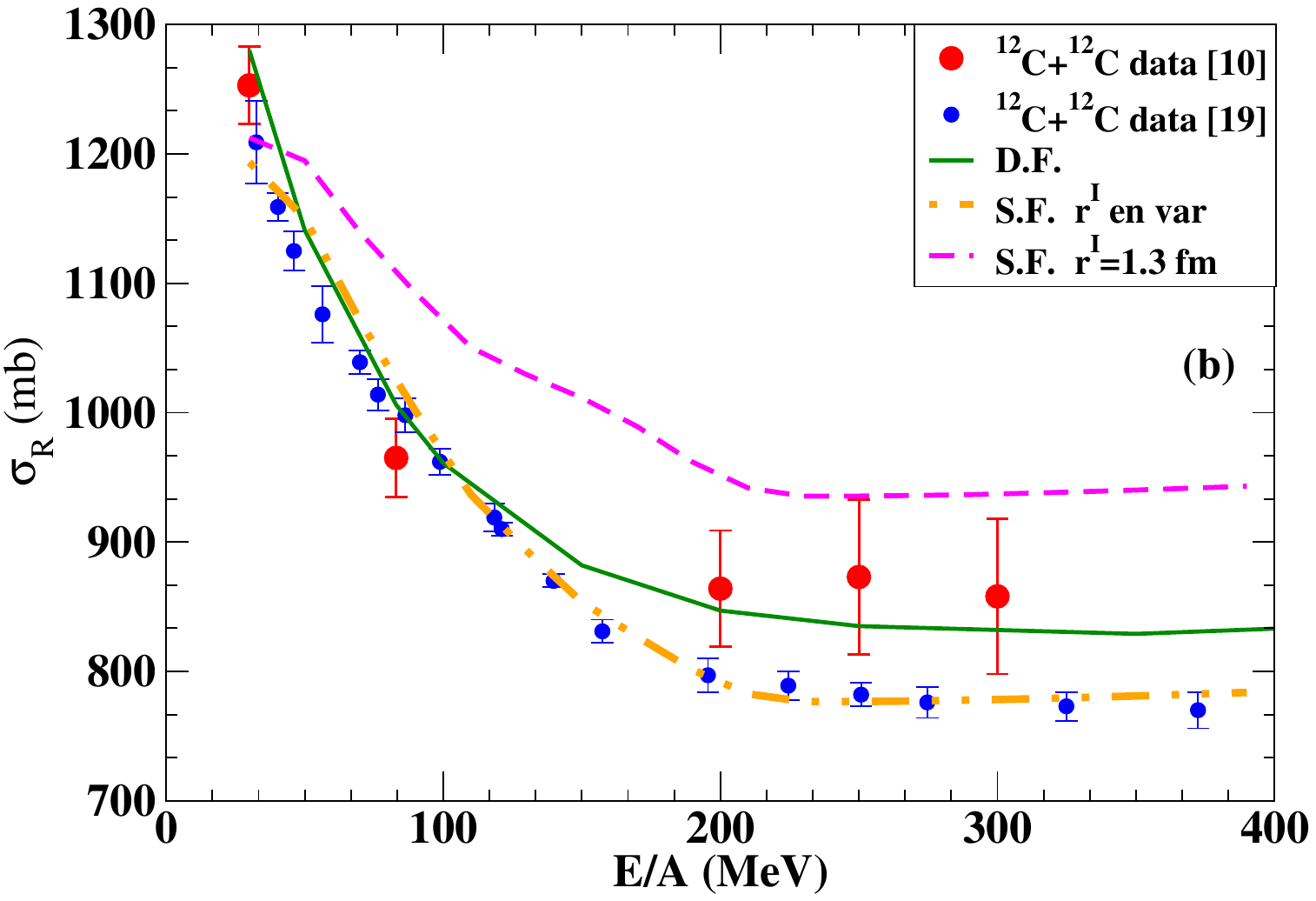}
\end{subfigure}

\caption  {Comparison of experimental reaction cross sections (circles with error bars) and theoretical values according to Eq.(6) within S.F.  and D.F. potentials (dot-dashed  and full lines  respectively), for the scattering of   $^{12}${C}+ $^{9}$Be (a) and $^{12}$C+ $^{12}$C (b). The magenta dashed lines  in both  panels represent the S.F. results obtained using  a fixed value $r^I=1.3 $ fm for the radius parameter of the imaginary phenomenological optical potential.   The dot-dashed lines correspond to an energy dependent $r^I$. according to Table \ref{tabri}. See text for details. Data points are from Ref. \cite{take}. In the lower panel the large red points are from Ref. \cite{kox}.}
\label{diffr0} 
\end{figure}

In Fig. \ref{pot4} a number of such imaginary potentials are shown for  the $^{12}$C-$^{12}$C system  at 83 and 300 MeV as indicated in the legend. We show D.F. potentials obtained with   the HF and no-core-shell-model   (NCSM) densities obtained from the nn4lo \cite{n4lo} interaction and   S.F. potentials obtained with the potentials of Table II, varying the $r^I$ values and the NCSM and HFB  densities. We  will see in the following that, in order to reproduce the experimental cross sections, the  $r^I$ parameter needs to be energy dependent when the $n$-T phenomenological potential is used to build up the $NN$ potential. The lower figure shows the potentials from the NCSM density at $300 A$ MeV,  where the D.F. has been renormalized by a factor 0.4 in order to compare it directly to the S.F. potential and to emphasize the difference in shape and rms radius. The D.F. potentials  shown in panel (a)  of Fig. \ref{pot4} are  deeper and have smaller rms  radii  than the S.F. potentials which are characterized instead by longer tails and larger rms values while their volume integrals are smaller than those of the D.F. potentials; see also Table \ref{TC_12C+12C}.  In the same table    the values of calculated reaction cross sections at 83 and $300A$ MeV are given. Incident energies  are indicated  in the  first column,  strong absorption radius parameters within the single and D.F. methods using the HFB densities are listed in the third column, while the  fourth column provides the volume integrals for active particles of the imaginary potentials. The next  columns contain the  theoretical cross sections calculated with various densities. On the left-hand side of each of them are the rms radii of the corresponding imaginary potentials shown in Fig. \ref{pot4}. Typically an increase of 5\% in the rms value results in a similar increase in the calculated reaction cross section, Eq.(6),  similarly to what we have noticed for the $n$-target potential. The values of Table \ref{TC_12C+12C} indicate that the volume integrals are the same for all densities, as they are normalized to the number of particles, while the rms values are different. However, they   obviously depend on the energy and on the method used to build the potential. On the other hand for each D.F. potential the rms values are independent of the energy because they are just determined by the densities.  This is consistent with the results of Ref. \cite{noi2}. The accuracy of our results can be discussed for example in comparison to Refs. \cite{Sakura1, Sakura100}. In that work the data for $^{12}$C+ $^{12}${C} elastic scattering were studied  at $100 A$ MeV using  microscopic coupled-channel calculations with the explicit goal to check the effect of repulsive three-body forces. The potential between the colliding
nuclei was determined by the double folding method with three different complex $g$-matrix interactions,  and also the reaction cross section was calculated. The  calculated value which agreed better with the data was $\sigma_R=950$ mb, obtained with the MPa interaction \cite{yama} and a renormalization factor $N_W=0.57$ for the imaginary potential. The  MPa interaction includes repulsive three-body forces.  It is interesting to note that with our S.F. potential we obtain 969 and 953 mb with the HFB and HF densities respectively, without any renormalization for the potential, while the experimental value is 962 mb. With the  D.F. potential and the HFB densities we obtain 980 mb. Also, similarly to what is shown in Fig. \ref{diffr0} and Table \ref{TC_12C+12C} for the D.F. and S.F. potentials at 300 MeV, we find that at 100 MeV the depth of the D.F. potential should be renormalized by a factor 0.4 with respect to the S.F. potential depths to make their values similar. However as noticed at 300 MeV, also at 100 MeV the rms radii would be very different, namely 3.75 and 3.43 fm for the S.F. and D.F. potentials respectively. This confirms the fact that a simple D.F. potential calculated according to Eqs. (\ref{4}) and (\ref{5}) would be far too absorptive because it does not contain  {\it in-medium } effects which instead are partially contained in the microscopic potential of Ref. \cite{Sakura100} thanks to the introduction of the three-body repulsive force. Thus such potentials need a  not too strong renormalization. In light of such microscopic method results, one possible interpretation for our surface dominated  $n$-T phenomenological potentials  which give rise to relatively shallow but "wide" $NN$ potentials, cf. Figs. \ref{xs4} and \ref{pot4}, is that they contain in a effective way the effects of short range repulsion pushing most  $nn$ interactions  to the surface. 

Another interesting comparison 
can be done with the MOL method of Ref. \cite{suz}, in particular their Eq.(10) for the $S$ matrix,  \begin{equation}\exp{\left(i\tilde \chi_{OLA}({\bf b})\right) }=\exp{\left(-\int d{\bf r} \rho_p({\bf r}) \Gamma_{NT}({\bf b+\xi})\right)},\label{9}\end{equation} 
contains the profile function
 \begin{equation} \Gamma_{NT}({\bf b})=\left( \sigma_{1}(1-i\alpha_1) {e^{-{\bf b}^2/{2\beta_1}}\over {4\pi\beta_1}}+\sigma_2(1-i\alpha_2){e^{-{\bf b}^2/{2\beta_2}}\over  {4\pi\beta_2}} \right), \label{10}\end{equation} 
 with $\sigma_{1,2}$ and $\beta_{1,2}$  given by the values in Table I of \cite{suz} and   $\rho_p$  given by Eq. (75) and Table 2 of  \cite{ogawa}.
 It could be interpreted as a S.F. model in which $\Gamma_\text{{NT}}$
would be  the result of the $z$-integration of an effective nucleon-target  potential of Gaussian shape  with imaginary part
\begin {equation}W_{MOL}({\bf r})={1\over 2} \hbar v \left( \sigma_{1}{ e^{-r^2/{2\beta_1}}\over ({2\pi\beta_1})^{3/2}}+\sigma_2 {e^{-r^2/{2\beta_2}}\over  ({2\pi\beta_2})^{3/2}} \right).\label{11}\end{equation} 
 Such a potential, shown in  Fig. \ref{xs4} by the blue line  for $n+ ^{12}$C, shows a repulsive behavior at very short distances, which could be interpreted as an effective representation of short distance repulsion originating  in the  three-body terms of the chiral interaction as used for example in the microscopic model of  \cite{Sakura100}. On the other hand in Fig. \ref{pot4}  the full magenta line with blue dots  shows the corresponding $NN$ imaginary potential for the system  $^{12}$C+$^{12}${C} at $300 A$ MeV. It has  a volume integral of 184 MeV fm$^3$  and rms radius 3.48 fm, consistent  with our S.F. results of Table \ref{TC_12C+12C}. In particular we notice the same large distance behavior as in  our best S.F. potential. Thus the  modifications to the MOL  parameters introduced in Ref. \cite{take}, which the authors mentioned  are not  easily interpreted from the physical point of view, might represent an effective way to obtain the correct energy and radial dependence of their "effective" NT imaginary potential.

From the discussion of our  results it appears that Hartree-Fock  and HFB densities are the best for  reproducing  the experimental reaction cross section values, and indeed they are used in most codes related to exotic nuclei reactions.  Besides the  system $^{12}$C+ $^{12}$C, using HFB densities we study  also the systems $^{9}$Be+ $^{12}$C,  $^{20}$Ne+ $^{12}$C, and $^{n}$Ca+ $^{12}$C.  The energies of the scattering and cross sections and other relevant parameters are given in Tables \ref{TC_12C+12C}, \ref{20Ne+12C}, and \ref{Ca+12C}. In particular  as a significative parameter we  provide also the {\it strong-absorption radius} $R_s$ \cite{bass, me1}, obtained from the S matrices  as the radius where $\mid S_\text{{PT}}(R_s)\mid ^2={1\over 2}$, and in particular the  "strong absorption radius parameter"  $r_s$  extracted from 
\begin{equation} R_s=r_s(E_{\text{inc}})(A_P^{1/3}+A_T^{1/3}).\label{rs}\end{equation}
The values of this parameter in  Tables \ref{TC_12C+12C}, \ref{20Ne+12C}, and \ref{Ca+12C} indicate  also that the S.F. potentials provide longer range absorption
than the D.F. potential. The cross sections and rms radii in Tables \ref{20Ne+12C}, and \ref{Ca+12C} were calculated with two different options for the $r^I$ parameter of the phenomenological imaginary potential. The values in parentheses were obtained with $r^I=1.3$ fm while the other values were obtained with the prescription of Table \ref{tabri}. The best agreement with the data is obtained with an energy dependent $r^I$, as we  discuss further in the following.

 Figure \ref{diffr0}  presents the energy dependence of the calculated and experimental reaction cross sections  \cite{kox,take} for  $^{9}$Be+ $^{12}$C and  $^{12}$C+ $^{12}$C. There are two curves showing  results    obtained within the S.F. model: one (dot-dashed line), using in the phenomenological imaginary part  of the $n$-T potential the  radius parameter $r^I$ which depends on the incident energy according to Table \ref{tabri}, provides the best agreement with the data while the other (dashed line) using the standard $r^I=1.3$ fm, corresponds to values larger than the data. This is consistent with the results  in Tables \ref{20Ne+12C}, and \ref{Ca+12C}. It is interesting that the small change in $r^I$ brings the S.F. results in much better agreement with the data. The full lines are  D.F. results which are in between the two S.F. curves. What we have found is interesting because it agrees with what has been discussed in other works like Ref. \cite{take}. Namely it shows that modifications might be necessary  in  reaction models when including ingredients which  successfully reproduce  simpler reactions. In the case of the D.F. model it is evident that not only is the idea of a $NN$ reaction being a collection of   $nn$ free reactions is too simple, but  so is  the S.F. description of a collection of free, independent nucleons interacting with a nucleus via  optical model potentials. However,  at the moment it seems that simple, understandable  modifications are sufficient to reproduce the data. For example, the reduction in the radius parameter found useful in our model might indicate that, when a nucleus scatters from another nucleus,  as the energy increases its nucleons interact  with those of the other nucleus at smaller distances than a free nucleon interacts with the nucleons of a nucleus. 
 
\section {Conclusions}

 In this paper we  obtained an excellent phenomenological $n$- $^{12}$C optical potential which fits the total cross sections up to 500 MeV. We  then single folded it with various projectile densities and  studied the systems $^{12}$C+ $^{12}$C, $^{9}$Be+ $^{12}$C, $^{20}$Ne+ $^{12}$C, and $^{n}$Ca+ $^{12}$C, finding that the energy dependence of the reaction cross section data can be fitted by introducing a simple energy dependence in  the radius parameter of the imaginary $n$-target potential. D.F potentials  were also  calculated and it was shown once again that they are too deep and too "narrow."  On the other hand we have shown that the MOL method to calculate phase shifts, in which nucleon-target multiple scattering effects are taken into account, would provide  potentials with characteristics similar to ours. The general conclusion of our study is then that it is necessary that the imaginary part of microscopic and/or semi-phenomenological optical potentials contains higher order and in-medium  effects. Also it would be useful to study further the importance of short range repulsion and/or or the effect of the  three-body  force which might be at the origin of the  necessary reduction of the strength of the potential at short distances. As a next step our S.F. method  could  be also tested by evaluating  the $S$ matrices that are  necessary in the eikonal formalism of nuclear breakup.\vspace{15pt}

\section*{Acknowledgements}
We are very grateful to  Mack Atckinson for providing us with the unpublished  calculations with the DOM potential shown in Fig.\ref{xs1}, to   Petr Navr\'atil and Michael Gennari for the numerical values of the NCSM densities, and to Carlotta Giusti and Matteo Vorabbi for  comments on the manuscript. One of us (I.M.) thanks  M. Gaidarov and colleagues  for allowing her to run and  use results from the  code  HFBTHO\cite{HFB}.


\begin{thebibliography}{30}

\bibitem{fesh}  H. Feshbach, Ann. Phys. (N. Y. ) 5, (1958) 357.

\bibitem{fesh1} H. Feshbach, Ann. Rev. Nucl. Sci., 8 (1958) 49.   

\bibitem{3} G.W. Greenlees, G.J. Pyle and Y.C. Tang, Phys. Rev. 171
(1968) 1115.
\bibitem{4} J.P. Vary and C.B. Dover, in Proceedings of the Second High
Energy Heavy-Ion Summer Study, Lawrence Berkeley  National Laboratory, July,
1974 (unpublished).
\bibitem{SL} G.R. Satchler and  W.G. Love, Phys. Rep. { 55} 183  (1979).
\bibitem{S2} G.R. Satchler, in Proceedings of La Rabida international Summer School on Heavy Ion Collisions,
La Rabida (Huelva), Spain,June 7-19, 1982 (unpublished). https://inis.iaea.org/search/search.aspx?orig\_q=RN:14722968. 

\bibitem {59} R. J. Glauber, in Lectures in Theoretical Physics, Vol. 1, edited by W. E. Brittin and L. G. Dunham (Interscience, New York, 1959), p. 315.
\bibitem{thesis} A. Bonaccorso, "A microscopic theory of the alpha-nucleus optical potential", Ph.D.  thesis, University of Oxford, 1980 (unpublished),  https://ora.ox.ac.uk/objects/uuid:d77df433-a09d-46c2-b94b-4d032fcf39b4
\bibitem {dvp} R.M. De Vries, J.C. Peng, Phys. Rev. C 22 (1980) 1055.



\bibitem{kox} S. Kox et al., Phys. Rev. C35,1678 (1987).

\bibitem{sfpang} Y. P. Xu and D. Y. Pang, Phys. Rev.  C 87, 044605 (2013).


\bibitem{jlm} J. P. Jeukenne, A. Lejeune, and C. Mahaux, Phys. Rev. C 16, 80
(1977).
\bibitem{jlm1}E. Bauge, J. P. Delaroche, and M. Girod,Phys. Rev. C 63, 024607
(2001).

\bibitem{jin} Y. Lu, J. Lei, and Z. Ren, Phys. Rev. C 108, 024612 (2023).
\bibitem{KD} A. J. Koning , J. P. Delaroche, Nucl. Phys. A 713
 231 (2003). 

\bibitem{suz} B. Abu-Ibrahim and Y. Suzuki,  Phys. Rev. C 62, 034608 (2000), Phys. Rev. C 61, 051601(R) (2000).

\bibitem{Huss} M.S.  Hussein, R. A. Rego, C. A. Bertulani, Phys. Rep.  201, (1991) 279Ñ334. 
\bibitem{carlos} C. A. Bertulani, and C. De Conti, Phys. Rev. C {81},   064603  (2010).
\bibitem{take} M. Takechi, M. Fukuda, M. Mihara, K. Tanaka, T. Chinda,
T. Matsumasa, M. Nishimoto, R. Matsumiya, Y. Nakashima, H. Matsubara, K. Matsuta, T. Minamisono, T. Ohtsubo, T. Izumikawa, S. Momota, T. Suzuki, T. Yamaguchi, R. Koyama, W. Shinozaki, M. Takahashi, A. Takizawa, T. Matsuyama, S. Nakajima, K. Kobayashi, M. Hosoi, T. Suda, M. Sasaki, S. Sato, M. Kanazawa, and A. Kitagawa, Phys. Rev. C 79, 061601(R) (2009).

\bibitem{fitbeta} D. T. Tran, H. J. Ong, T. T. Nguyen, I. Tanihata, N. Aoi, Y. Ayyad, P. Y. Chan, M. Fukuda, T. Hashimoto, T. H. Hoang, E. Ideguchi, A. Inoue, T. Kawabata, L. H. Khiem, W. P. Lin, K. Matsuta, M. Mihara, S. Momota, D. Nagae, N. D. Nguyen, D. Nishimura, A. Ozawa, P. P. Ren, H. Sakaguchi, J. Tanaka, M. Takechi, S. Terashima, R. Wada, and T. Yamamoto, Phys. Rev. C 94, 064604 (2016).

\bibitem{bobwim}W.H. Dickhoff and R.J. Charity, Prog.Part.  Nucl. Phys. 105  252 (2019).
\bibitem{ch}  M. Burrows, C. Elster, S. P. Weppner, K. D. Launey, P. Maris, A. Nogga, and G. Popa, Phys. Rev. C 99, 044603 (2019).
\bibitem{carlo} A. Idini, C.Barbieri, P. Navr\'atil, Phys. Rev. Lett. 123, 092501 (2019).
\bibitem{Vorabbi1} M. Vorabbi, M.  Gennari, P. Finelli, C. Giusti, P. Navr\'atil, and R. Machleidt, Phys. Rev. C103, 024604 (2021).
\bibitem{n4lo} D. R. Entem, R. Machleidt, and Y. Nosyk, Phys. Rev. C 96,
024004 (2017).
\bibitem{Vorabbi} P. Finelli, M. Vorabbi, C.Giusti, J. Phys.: Conf. Ser. 2453, 012026 (2023), and refrences therein.

\bibitem{furu} T. Furumoto, K. Tsubakihara, S. Ebata, W. Horiuchi, Phys. Rev. C 99, 034605 (2019).
\bibitem{Sakura} T. Furumoto, Y. Sakuragi, and Y. Yamamoto
Phys. Rev. C 78, 044610 (2008)
\bibitem{Sakura1} T. Furumoto, W. Horiuchi, M. Takashina, Y. Yamamoto, and Y. Sakuragi
Phys. Rev. C 85, 044607 (2012). 
\bibitem{Sakura100} Qu,W.W., Zhang,G.L., Terashima,S., Furumoto,T., Ayyad,Y., Chen,Z.Q., Guo,C.L., Inoue,A., Le,X.Y., Ong,H.J., Pang,D.Y., Sakaguchi,H., Sakuragi,Y., Sun,B.H., Tamii,A., Tanihata,I., Wang,T.F., Wada,R., Yamamoto,Y.,  Phys Rev. C95, 044616 (2017) and references therein.
\bibitem{ogata} M. Toyokawa, M. Yahiro, T. Matsumoto, K. Minomo, K. Ogata, and M. Kohno, Phys. Rev. C92, 024618 (2015) and references therein.


\bibitem{bobme} A. Bonaccorso and R. J. Charity,  Phys. Rev.  C { 89},  024619   (2014).
\bibitem{MS} C. Mahaux and R. Sartor, Adv. Nucl. Phys. 20, 1 (1991).
\bibitem{noi1} A. Bonaccorso, F. Carstoiu, R. J. Charity, R. Kumar and G. Salvioni, Few-Body Syst.  { 57},  331 (2016).
\bibitem{noi2} A. Bonaccorso, F. Carstoiu, R. J. Charity, Phys. Rev. C  {94},  034604 (2016).
\bibitem{imane} Imane Moumene and Angela Bonaccorso, Nucl. Phys. {A1006} 122109 (2021).
\bibitem{GSI}D. Boscolo et al., Front. Oncol. 11, 737050 (2021)  https://doi.org/10.3389/fonc.2021.737050

\bibitem{filo} C. Hebborn, T. R. Whitehead, A. E. Lovell, and F. M. Nunes, Phys. Rev. C 108, 014601 (2023).

\bibitem{luoni} F Luoni, F Horst, C A Reidel, A Quarz, L Bagnale, L Sihver, U Weber, R B Norman , W de Wet and M Giraudo, G Santin, J W Norbury and M Durante, New Journal of Physics  10, 101201(2021).
https://dx.doi.org/10.1088/1367-2630/ac27e1
\bibitem{pet} Malouff TD, Mahajan A, Krishnan S,
Beltran C, Seneviratne DS and
Trifiletti DM Front. Oncol. 10:82. doi: 10.3389/fonc.2020.00082
https://kcch.kanagawa-pho.jp/i-rock/english/medical/
\bibitem{kuni}S. Kunieda et al., Eur. Phys. J. A 59, 2 (2023). 

\bibitem{tanih1} I. Tanihata et al., Phys. Letters B 160 (1985) 380.
\bibitem {hor2} B. Abu-Ibrahim, W. Horiuchi, A. Kohama, and Y. Suzuki, Phys. Rev. C 77, 034607 (2008)

\bibitem{Ca+12C} M. Tanaka et al. Phys. Rev. Lett. 124, 102501 (2020).
\bibitem {hor1} W. Horiuchi, Y. Suzuki, B. Abu-Ibrahim, and A. Kohama, Phys. Rev.  C 75, 044607 (2007).
\bibitem{ozawa} A. Ozawa et al., Nucl. Phys. A { 691}, 599 (2001).
A. Ozawa,  AIP Conf. Proc. {865}, 57 (2006); http://dx.doi.org/10.1063/1.2398828 

\bibitem{tanih} Isao Tanihata, Herve Savajols, Rituparna Kanungo, Prog. Part. Nucl. Phys. 68 (2013) 215, and references therein.

\bibitem{AB0} Angela Bonaccorso, Progress in Particle and Nuclear Physics, 101(2018)  1-54, and references therein.

\bibitem{exf} EXFOR nuclear data library [http://www.nds.
iaea.org/exfor/exfor.htm].


\bibitem{mack} M. Atckinson, (private communication).
\bibitem{jinme} Jin Lei, A. Bonaccorso, Phys. Lett. B  813, 136032 (2021).



\bibitem{Wiringa}  R. B. Wiringa, R. Schiavilla, S. C. Pieper, and J. Carlson,
Phys. Rev. C 89, 024305 (2014); M. Piarulli, S. Pastore, R. B. Wiringa, S. Brusilow, and R. Lim, ibid. 107, 014314 (2023), https://www.phy.anl.gov/theory/research/density/, and references therein.
\bibitem{nav} V. Som\`{a}, P. Navr\'{a}til, F. Raimondi, C. Barbieri, and T. Duguet, Phys. Rev. C 101, 014318 (2020).
\bibitem{HFB}M. V. Stoitsov, N. Schunck, M. Kortelainen, N. Michel, H. Nam, E. Olsen, J. Sarich, and S. Wild, Comput. Phys. Commun. 184, 1592 (2013); M. V. Stoitsov, J. Dobaczewski, W. Nazarewicz, and P. Ring, ibid. 167, 43 (2005).
\bibitem{SkM*} J. Bartel, E. Quentin, M. Brack, C. Guet and H.-B. Hakansson, Nucl. Phys. A 386 (1982) 79. 

\bibitem{sch} P. Schwaller et al., Nuclei. Phys. A316 317(1979).

\bibitem{yama} Y. Yamamoto, T. Furumoto, N. Yasutake, and Th.A. Rijken,
1 Eur. Phys. J. A (2016) 52: 19

\bibitem{ogawa} Y. Ogawa, K. Yabana,  and Y. Suzuki, Nucl. Phys. A543 722 (1992).
\bibitem{bass} R. Bass, {\it Nuclear Reactions with Heavy Ions}, Springer-Verlag, Berlin, Heidelberg, New York, 1980, Sec. 3.3.
\bibitem{me1} A. Bonaccorso, D. M. Brink and L. Lo Monaco, J. Phys. G {13} 1407 (1987).













  \end{thebibliography}
\end{document}